\documentclass[a4paper,11pt]{article}

\usepackage{pos}
\usepackage{graphicx}
\usepackage{caption}

\usepackage{blindtext}
\usepackage[v1compatible]{axodraw2}
\usepackage{microtype}

\usepackage{hyperref}

\usepackage{ulem}

\graphicspath{{pictures}}

\newcommand{\valencia}{
Instituto de F\'{\i}sica Corpuscular,\\
Universitat de Val\`{e}ncia,\\ 
Consejo Superior de Investigaciones Cient\'{\i}ficas, \\ Parc Cient\'{\i}fic, E-46980 Paterna, Valencia, Spain.
}

\title{Grover's Quantum Search Algorithm of Causal Multiloop Feynman Integrals.}
\ShortTitle{Quantum Algorithm for Feynman Integrals}

\author*[a]{Andrés E. Rentería-Olivo}

\affiliation[a]{\valencia
}

\emailAdd{andres.renteria@ific.uv.es}

\abstract{
A proof-of-concept application of a quantum algorithm to multiloop Feynman integrals in the 
Loop-Tree Duality (LTD) framework is applied to a representative four-loop topology. Bootstrapping
causality in the LTD formalism, is a suitable problem to address with quantum computers 
given the straightforward possibility to  encode the two on-shell states of a propagator on the two states of a qubit. 
A modification of Grover's quantum search algorithm is developed and the quantum algorithm is successfully
implemented on IBM Quantum and QUTE simulators. 
}

\FullConference{%
    41st International Conference on High Energy physics - ICHEP2022\\
    6-13 July, 2022\\
    Bologna, Italy
}

\def\qon#1{q_{#1,0}^{(+)}}

\def\ket#1{|{#1}\rangle}

\begin{document}
\maketitle
\renewcommand{\thefootnote}{\fnsymbol{footnote}}

\section{Introduction}
High-precision theoretical predictions for current and future high-energy particle colliders require novel techniques 
to deal with the higher orders in perturbation theory of scattering amplitudes. In this context, the loop-tree duality (LTD) 
\cite{Catani:2008xa,Aguilera-Verdugo:2020set,deJesusAguilera-Verdugo:2021mvg}
formalism exhibits interesting mathematical properties of high potential to overcoming current limitations.
A remarkable property of LTD is the possibility of representing the causal nature of Feynman diagrams and scattering 
amplitudes, leading to an intuitive understanding of the singular structure of loop integrals.
This manifestly causal representation~\cite{Aguilera-Verdugo:2020set,Aguilera-Verdugo:2020kzc,Ramirez-Uribe:2020hes} allows to 
replace the original multiloop Feynman diagrams into a class of multiloop topologies 
defined by collapsing propagators into edges where the two on-shell states of propagators are naturally encoded 
by the two states of a qubit, leading to explore the application of quantum algorithms,
for instance, a Grover's quantum search algorithm~\cite{Ramirez-Uribe:2021ubp} and a Variational Quantum Eigensolver~\cite{Clemente:2022nll} approach.
In this work we focus on the former approach.

\setlength{\belowdisplayskip}{5pt} \setlength{\belowdisplayshortskip}{5pt}
\setlength{\abovedisplayskip}{5pt} \setlength{\abovedisplayshortskip}{5pt}

\section{Loop-Tree Duality and Causality}
The causal representation of scattering amplitudes in the LTD formalism is obtained through the calculation of nested residues. 
In Refs.~\cite{deJesusAguilera-Verdugo:2021mvg,Aguilera-Verdugo:2020set} it is shown that scattering amplitudes can be written as
\begin{equation}
    {\cal A}_D^{(L)} = \int_{\vec \ell_1 \ldots \vec \ell_L} 
    \frac{1}{x_n} \sum_{\sigma  \in \Sigma} {\cal N}_{\sigma}\, \prod_{i=1}^{n-L} \, \frac{1}{\lambda_{\sigma(i)}^{h_{\sigma(i)}}} \
    + (\lambda^+ \leftrightarrow \lambda^-)~,
\label{eq:CausalRep}
\end{equation}
with $x_n = \prod_i 2\qon{i}$, $h_{\sigma(i)} = \pm 1$, $\cal{N}_{\sigma}$ a numerator determined by the interaction vertices of a specific theory and $\int_{\vec{\ell_s}} = -\mu^{4-d}(2\pi)^{1-d}\int{\rm d}^{d-1}\ell_s$,
the integration measure in the loop three-momentum space.
The Eq. (\ref{eq:CausalRep}) only involves denominators with positive on-shell energies 
$q_{i,0}^{(+)}=(\vec{q}_i^{\,2}+m_i^2-\imath0)^{1/2}$, 
added together in same-sign combinations in the so-called causal propagators, $1/\lambda_{\sigma(i)}^\pm$, with
\begin{equation}
    \lambda_{\sigma(i)}^\pm \equiv \lambda_p^\pm = \sum_{i\in p} \qon{i} \pm k_{p,0}~,
\end{equation}
where $\sigma(i)$ stands for a partition $p$ of the set of on-shell energies and the orientation of the energy components of 
the external momenta, $k_{p,0}$.
The causal structure of $\lambda_p^{\pm}$ is defined by the sign of $k_{p,0}$ when the propagators in the partition $p$ are set on-shell.
Each causal propagator is in a one-to-one correspondence with any possible threshold singularity of the amplitude, which contains overlapped thresholds that are known as causal entangled thresholds.
The combinations of entangled causal propagators represent causal thresholds that can occur simultaneously, which are collected 
in the set~$\Sigma$.

\section{Causal query of multiloop Feynman integrals: the four-eloop case}
\begin{figure}
\begin{minipage}{0.24\textwidth}\centering
\begin{axopicture}(50,50)
    \SetWidth{1} 
    \Vertex(    30      ,   30      ){2}
    \Vertex( 51.2133 , 51.2133 ){2}
    \Vertex( 51.2133 , 8.7867 ){2} 
    \Vertex( 8.7867 , 8.7867 ){2}
    \Vertex( 8.7867 , 51.2133 ){2}
    \Arc[arrow](30,30)( 30, 45,135)
    \Arc[arrow](30,30)( 30,135,225)
    \Arc[arrow](30,30)( 30,225,315)
    \Arc[arrow](30,30)( 30,315, 45)
    \Line[arrow,arrowpos=0.45,flip](30,30)( 51.2133 , 51.2133 )
    \Line[arrow,arrowpos=0.45,flip](30,30)( 51.2133 , 8.7867 )
    \Line[arrow,arrowpos=0.45,flip](30,30)( 8.7867 , 8.7867 )
    \Line[arrow,arrowpos=0.45,flip](30,30)( 8.7867 , 51.2133 )
    \Text( -5,27  )[]{\scriptsize$e_0$}
    \Text( 28,65  )[]{\scriptsize$e_1$}
    \Text( 66,32  )[]{\scriptsize$e_2$}
    \Text( 33,-5.5)[]{\scriptsize$e_3$}
    \Text( 15,25  )[]{\scriptsize$e_4$}
    \Text( 25,45  )[]{\scriptsize$e_5$}
    \Text( 45,35  )[]{\scriptsize$e_6$}
    \Text( 35,15  )[]{\scriptsize$e_7$}
\end{axopicture}
\end{minipage}
\begin{minipage}{0.749\textwidth}\centering
\includegraphics[width=0.8\textwidth]{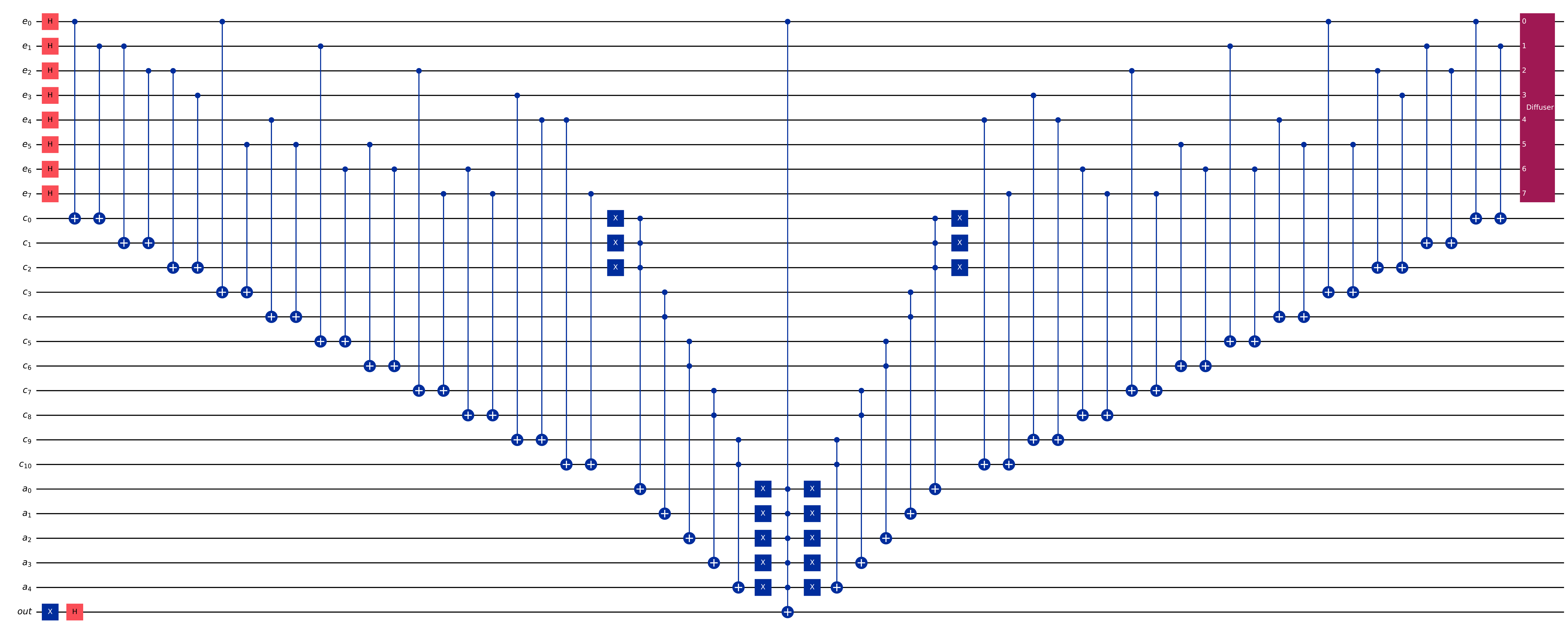}
\end{minipage}
\caption{
Left: Four-eloop topology with a fixed conventional orientation. 
Right: corresponding quantum circuit for the four-eloop topology.}
\label{fig:topolofyandCircuit}
\end{figure}

We develop a modified version~\cite{Ramirez-Uribe:2021ubp} of Grover's algorithm~\cite{Grover:1997fa} for querying 
the causal configurations from the LTD framework. We take as example the four-eloop topology, which consists in four loops made of edges (eloops)
with a central four-particle interaction vertex and we implement it on IBM Quantum\footnote{\href{https://quantum-computing.ibm.com/}{https://quantum-computing.ibm.com/}}.
The implementation of this quantum algorithm requires four registers, namely, $\ket{e}$, $\ket{c}$, $\ket{a}$ and $\ket{\it out}$. 
In the first register, $\ket{e}$, we encode the state of the $N=2^n$ edges of the topology and initialise it with a uniform superposition by
applying Hadamard gates, $\ket{e}=H^{\otimes n}\ket{0}$. 
The second register, $\ket{c}$, encodes the comparison 
between two adjacent edges $e_i$ and $e_j$ in binary Boolean clauses defined by
\begin{align}
    c_{ij} \equiv (e_i = e_j) \qquad \text{and} \qquad \bar{c}_{ij} \equiv (e_i \neq e_j) ~.
\end{align}
Each clause $\bar{c}_{ij}$ is implemented through two CNOT gates, each taking as control the corresponding qubits $e_i$ and $e_j$ respectively, and both taking as target the same qubit in the register $\ket{c}$. The clause $c_{ij}$ requires an extra NOT gate on the target qubit on the register $\ket{c}$.   
The third register, $\ket{a}$, stores the eloop clauses that probe the adjacent edges that compose a cyclic circuit within the diagram.
Specifically for the four-eloop topology, taking the conventional orientation of the edges in Fig. (\ref{fig:topolofyandCircuit}), the clauses are given by
\begin{align} 
      a_0^{(4)} &= \neg (c_{01} \wedge c_{12} \wedge c_{23}) ~, &
      a_1^{(4)} &= \neg (\bar{c}_{05} \wedge \bar{c}_{45})   ~, &
      a_2^{(4)} &= \neg (\bar{c}_{16} \wedge \bar{c}_{56})   ~, \nonumber \\
      a_3^{(4)} &= \neg (\bar{c}_{27} \wedge \bar{c}_{67})   ~, &
      a_4^{(4)} &= \neg (\bar{c}_{34} \wedge \bar{c}_{47})   ~. 
\end{align}
Each $a_i^{(4)}$ requires a multicontrolled Toffoli gate that takes as control its specific qubits $\ket{c_{ij}}$ and as target 
a qubit in the register $\ket{a}$ followed by a NOT gate.
The last register, $\ket{\it out}$, requires a single qubit initialised in the $\ket{-}=HX\ket{0}$ and is used as the Grover's marker. 
It stores the output of the quantum algorithm and is implemented through a multicontrolled Toffoli gate taking as control all the qubits 
from the register $\ket{a}$ and, if required, a qubit from the register $\ket{e}$.
The oracle is defined as 
\begin{align}
    U_w \ket{e} \ket{c} \ket{a} \ket{\it out} = (-1)^{f(a,e)} \ket{e} \ket{c} \ket{a} \ket{\it out} ~.
\end{align}
When the causal conditions are satisfied, $f(a,e)=1$, and marks the corresponding states; otherwise, 
if $f(a,e)=0$, they are left unchanged. 
For the four-eloop topology, the required Boolean marker is given by
\begin{align}
    f^{(4)}(a,e) = \left( ~ \bigwedge_{i=0}^4 a_i^{(4)} ~ \right) \wedge e_0 ~.
\end{align}
After storing the marked states in $\ket{\it out}$, the registers $\ket{a}$ and $\ket{c}$ are rotated back to the state $\ket{0}$ by 
applying the oracle operations in inverse order. The last step in the algorithm is the amplification of the marked states by applying 
the diffuser opertator on the register $\ket{e}$. We use the diffuser defined 
on IBM Quantum\footnote{\href{https://qiskit.org/textbook/ch-algorithms/grover.html}{https://qiskit.org}}.

\section{Conclusions}
The quantum algorithm developed in this work allowed us to successfully identify the causal singular configurations of the 
four-eloop topology in the LTD framework.
The algorithm was successfully implemented in IBM Quantum and 
QUTE simulators\footnote{\href{https://qute.ctic.es/hub/login}{https://qute.ctic.es}}. The result of the quantum algorithm is used to bootstrap the causal representation in LTD of representative multiloop topologies, allowing us a better understanding of this approach and enabling us to explore other techniques, such as a Variation Quantum approach, 
This challenging problem is closely related to the 
identification of directed acyclic graphs in graph theory, which can find application in other fields beyond particle physics.

\section{Acknowledgements}
\noindent
I would like to thank the team that collaborated in the development of this work, G. Rodrigo, S. Ramírez-Uribe, G. F. R. Sborlini
and L. Vale-Silva. To IBM Quantum for the access to IBMQ and also to Fundación Centro Tecnológico de la Información y Comunicación (CTIC) for the access to Quantum Testbed (QUTE).
This work is supported by the PRE2018-085925 grant from the Consejo Superior de Investigaciones Científicas (CSIC) of the Spanish Government, 
MCIN/AEI/10.13039/501100011033, and Grant No. PID2020-114473GB-I00.

\bibliographystyle{JHEP}

\end{document}